\documentclass{DISproc}

\newcommand{\jprBase}        {Phys.\ Rev.\xspace}
\newcommand{\jprd}      [1]  {\jprBase\ D~{\bf #1}}
\newcommand{\plb}       [1]  {\jplBase\ B~{\bf #1}}
\newcommand{\jplBase}        {Phys.\ Lett.\xspace}

\def\CP{\ensuremath{C\!P}\xspace}

\def\be{\begin{equation}}
\def\ee{\end{equation}}
\def\bea{\begin{eqnarray}}
\def\eea{\end{eqnarray}}

\newcommand{\btoptl}{\ensuremath{B \to \pi \tau \ell}}
\def\KS    {\ensuremath{K^0_{\scriptscriptstyle S}}\xspace}
\def\KL    {\ensuremath{K^0_{\scriptscriptstyle L}}\xspace}

\def\taum       {\ensuremath{\tau^-}\xspace}

\usepackage{relsize}
\def\babar{\mbox{\slshape B\kern-0.1em{\smaller A}\kern-0.1em
    B\kern-0.1em{\smaller A\kern-0.2em R}}}

\def\piz   {\ensuremath{\pi^0}\xspace}

\def\Bbar    {\kern 0.18em\overline{\kern -0.18em B}{}\xspace}

\def\BB      {\ensuremath{B\Bbar}\xspace} 
\def\Bz      {\ensuremath{B^0}\xspace}
\def\Bzb     {\ensuremath{\Bbar^0}\xspace}
\def\BzBzb   {\ensuremath{\Bz {\kern -0.16em \Bzb}}\xspace}
\def\Bu      {\ensuremath{B^+}\xspace}
\def\Bub     {\ensuremath{B^-}\xspace}
\def\Bp      {\ensuremath{\Bu}\xspace}

\def\Bpm     {\ensuremath{B^\pm}\xspace}

\def\BpBm    {\ensuremath{\Bu {\kern -0.16em \Bub}}\xspace}

\def\BB      {\ensuremath{B\Bbar}\xspace} 
\newcommand{\gevcc}{\ensuremath{{\mathrm{\,Ge\kern -0.1em V\!/}c^2}}\xspace}
\newcommand{\mevcc}{\ensuremath{{\mathrm{\,Me\kern -0.1em V\!/}c^2}}\xspace}
\def\mes       {\mbox{$m_{\rm ES}$}\xspace}
\def\jpsi     {\ensuremath{{J\mskip -3mu/\mskip -2mu\psi\mskip 2mu}}\xspace}

\def\nunub      {\ensuremath{\nu{\overline{\nu}}}\xspace}
\def\nub        {\ensuremath{\overline{\nu}}\xspace}
\def\nut        {\ensuremath{\nu_\tau}\xspace}
\def\nutb       {\ensuremath{\nub_\tau}\xspace}
\newcommand{\tautoenunu}{\ensuremath{\tau \to e \nunub}\xspace}
\newcommand{\tautomununu}{\ensuremath{\tau \to \mu \nunub}\xspace}
\newcommand{\tautopinu}{\ensuremath{\tau \to (n\pi^0) \pi \nu}\xspace}
\def\Dbar    {\kern 0.2em\overline{\kern -0.2em D}{}\xspace}
\def\taupksb     {\ensuremath{\tau^+\rightarrow\pi^+\KS\ \nutb}\xspace }
\def\taupks      {\ensuremath{\tau^-\rightarrow\pi^-\KS\,\nut}\xspace }
\def\asy         {\ensuremath{A_Q}\xspace }
\def\taupikspiz {\ensuremath{\taum \to \KS (\ge 0 \piz) \nut}}

\mathchardef\Upsilon="7107
\def\Y#1S{\ensuremath{\Upsilon{(#1S)}}\xspace}

\def\FourS {\Y4S}

\def\barn{\ensuremath{\rm \,b}\xspace}

\def\invpb {\ensuremath{\mbox{\,pb}^{-1}}\xspace}

\def\invfb   {\ensuremath{\mbox{\,fb}^{-1}}\xspace}

\def\ellp       {\ensuremath{\ell^+}\xspace}

\def\Bhll       {\ensuremath{\Bp\to h^- \ellp\ellp}\xspace}

\begin{document}
\title{Beyond Standard Model searches through heavy flavors at \babar}

\author{{\slshape Marco Bomben$^1$, on behalf of the BaBar collaboration}\\[1ex]
$^1$LPNHE - Barre 12-22, 1er Žtage - 4 place Jussieu - 75252 PARIS CEDEX 05}

\contribID{xy}

\doi  

\maketitle

\begin{abstract}
The \babar\ experiment recorded 471$\times10^6$ \BB pairs at the \FourS resonance (corresponding to an integrated luminosity of 429 \invfb). 
We present here a selection of recent results from the \babar\ collaboration: search for lepton-number violation in the decay $\Bp\to h^- \ellp\ellp$, search 
for lepton-flavor violation in $\Bpm\to h^{\pm}\tau\ell$ and \CP-violation in \taupikspiz.
\end{abstract}

\section{Search  for lepton-number violation in the decay \Bhll}

In the Standard Model (SM), the lepton number $L$ is conserved in low-energy collisions and decays.
Through neutrino oscillation $L$-conservation can be violated, for example in the process \Bhll; highly suppressed in the SM, this 
lepton-number violating process can be enhanced in several Beyond SM scenarios, such as left-right symmetric gauge theories, SO(10) supersymmetry, $R$-parity violating models, or extra-dimensions.

\begin{figure}[!htb]
\begin{center}
\includegraphics[width=5.5cm]{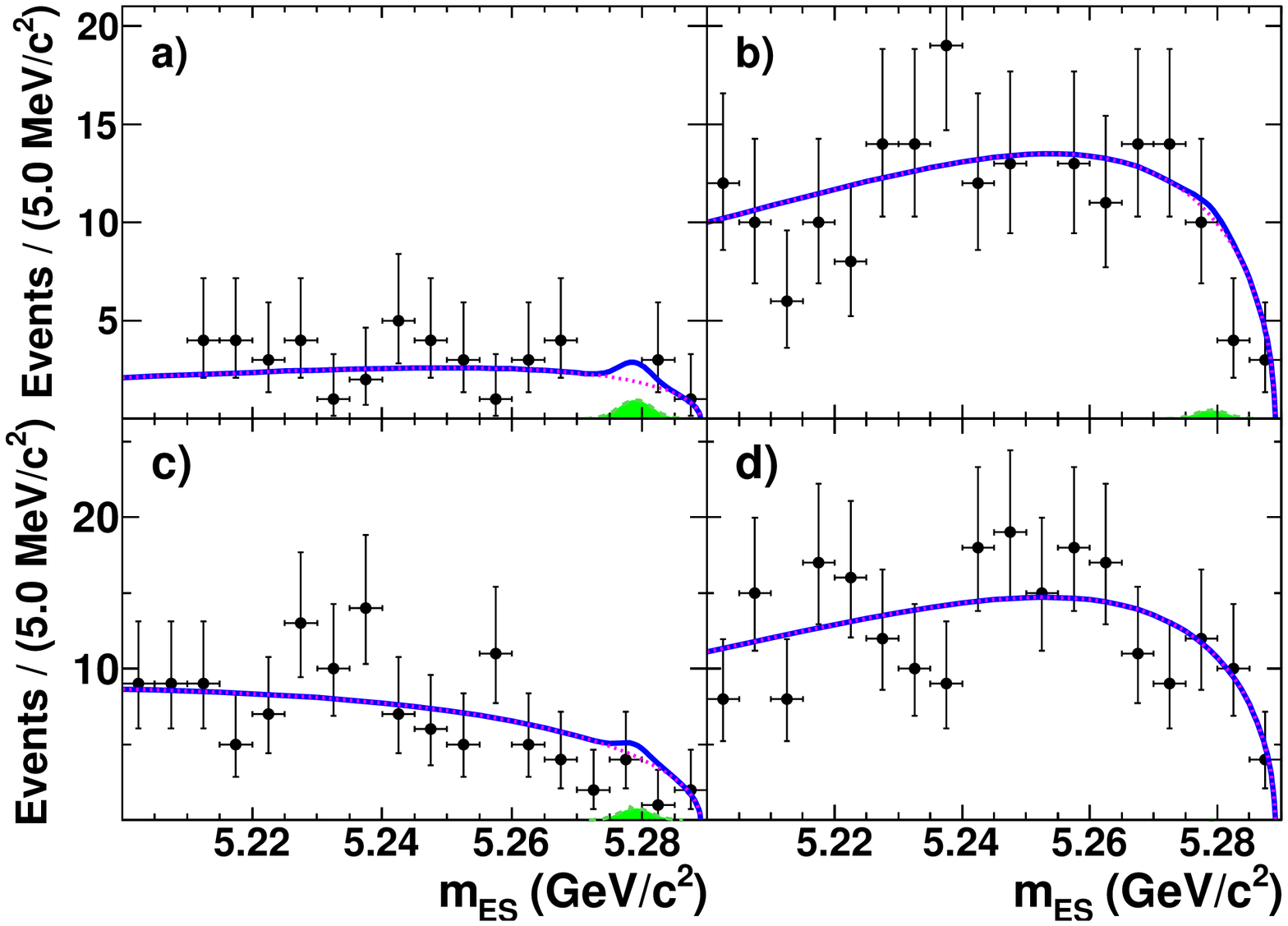}
\includegraphics[width=5.5cm]{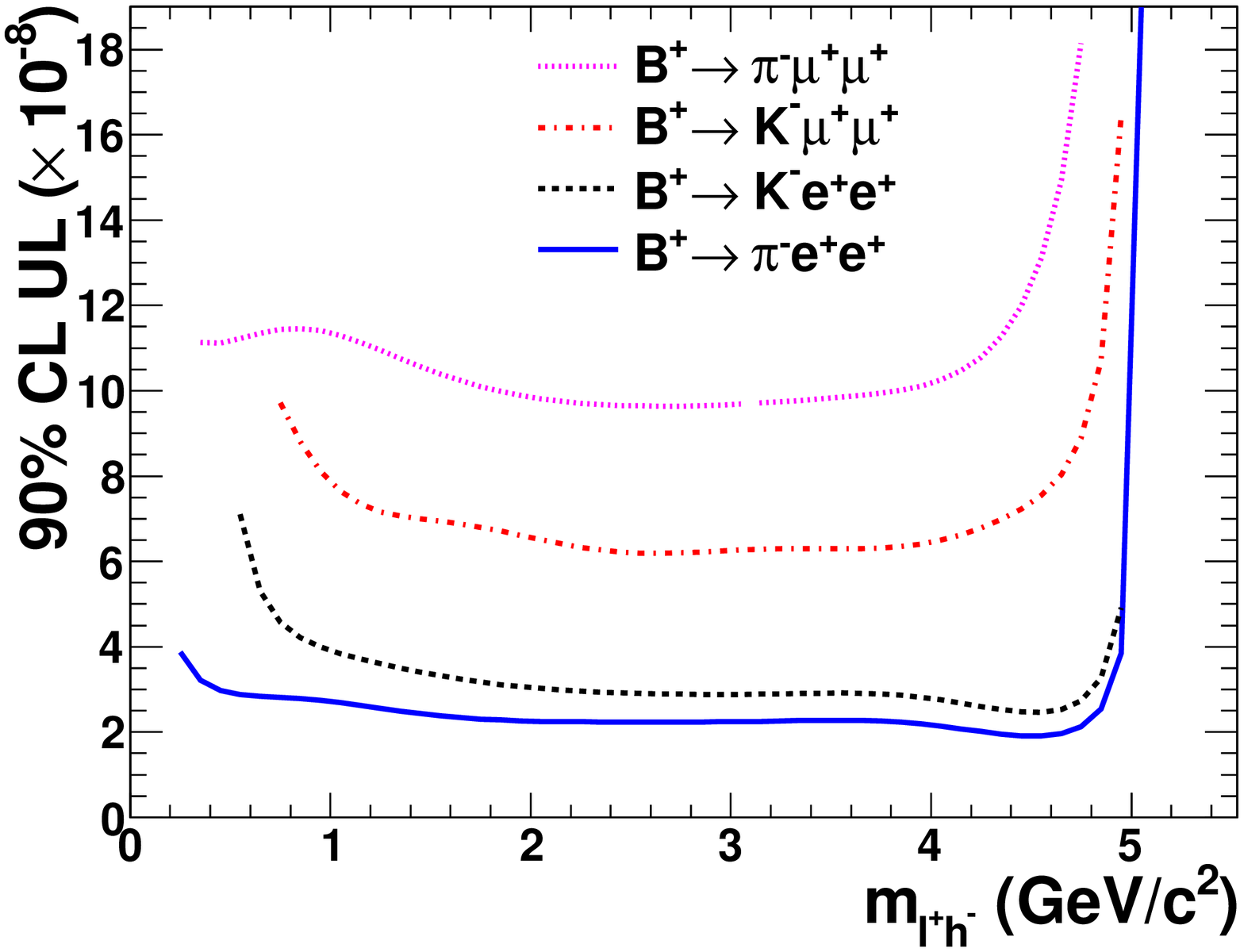}
\caption{Left: \mes\ distributions for a) $B^+ \to K^- e^+ e^+$, b) $B^+ \to K^- \mu^+ \mu^+$, c) $B^+ \to \pi^- e^+ e^+$, and d) $B^+ \to \pi^- \mu^+ \mu^+$.
The solid blue line is the total fit, the dotted magenta line is
the background, the solid green histogram is the signal. Right: 90\% CL upper limits on the branching fractions as a function of the mass $m_{\ell^+ h^-}$.}
\label{fig:hll}
\end{center}
\end{figure}

In this analysis~\cite{ref:lv}, four final states are considered: the decay channel is $B^+ \to h^- \ell^+ \ell^+$ with $h=K,\pi$ and $l=e,\mu$.
 The continuum and $B\Bbar$ backgrounds are rejected using bagged decision trees (BDT), based on $\Delta E$ (the difference between the expected $B$ energy and the reconstructed $B$ energy), event shape and vertexing variables. A likelihood ratio $R$ is constructed from the $B\Bbar$ BDT.  
Unbinned maximum likelihood fits of \mes\ and $R$, the likelihood ratio, are performed for each of the four modes. We use the $B^+ \to \jpsi h^+$ data control sample to obtain the \mes\ fit parameters. The fits are shown in Fig.~\ref{fig:hll}, where we observe that no signal is seen. From these negative results, we obtain upper limits at 90\% confidence level (CL) on the four channels: $\mathcal{B}(B^+ \to K^- e^+ e^+)<3.0\times10^{-8}$, $\mathcal{B}(B^+ \to K^- \mu^+ \mu^+)<6.7\times10^{-8}$, $\mathcal{B}(B^+ \to \pi^- e^+ e^+)<2.3\times10^{-8}$, and $\mathcal{B}(B^+ \to \pi^- \mu^+ \mu^+)<10.7\times10^{-8}$. The upper limits for the $h e^+ e^+$ channels are 40-70 times more stringent than previous limits set by other experiments.

\section{A search for the decay modes $B^\pm \to h^\pm \tau \ell$}

FCNC and charged lepton flavor violation are forbidden in the SM at tree level. However, in many extensions of the SM, these effects could be enhanced, especially for the second and third generation~\cite{ref:htltheo}. We study~\cite{ref:htl} the eight final states $B^\pm \to h^\pm \tau \ell$, with $h=K,\pi$ and $\ell = e, \mu$. The final states $B^\pm \to K^\pm \tau e$, $B^\pm \to \pi^\pm \tau \mu$, and $B^\pm \to \pi^\pm \tau e$ have never been done before. 
We fully reconstruct the hadronic $B$ on one side (the ``tag'' $B$) using final states of the type $B^- \to D^{(*)0}X^-$, where $X^-$ is composed of $\pi^\pm$, $K^\pm$, \KS, and $\pi^0$. This determines the three-momentum of the other $B$ (the ``signal'' $B$) on the other side and thus allows us to indirectly reconstruct the $\tau$ lepton through:
\begin{equation*}
\vec p_\tau  =  -\vec p_{\rm tag} - \vec p_h - \vec p_\ell; \;\;
E_\tau  =  E_{\rm beam} - E_h - E_\ell; \;\;
m_\tau  =  \sqrt{ E_\tau^2 - |\vec{p}_\tau|^2 },
\end{equation*}
where ($E_\tau$, $\vec p_\tau$), ($E_h$, $\vec p_h$), and ($E_\ell$, $\vec p_\ell$)
are the corresponding four-momenta of the reconstructed signal objects, and where $\vec p_{\rm tag}$ is the three-momentum of the tag $B$, and $E_{\rm beam}$ the beam energy. The $\tau$ is required to decay to a ``one-prong'' final state: $\tautoenunu$, $\tautomununu$, and $\tautopinu$ with $n \ge 0$. The signal branching fraction is determined by using the ratio of the number of $B^\pm \to h^\pm \tau \ell$ signal candidates to the yield of control samples of $B^+ \to \Dbar^{(*)0} \ell^+ \nu; \Dbar^0 \to K^+ \pi^-$ events from a fully reconstructed hadronic $B^\pm$ decay sample.

The background is mainly coming from semileptonic $B$ decays  and from semileptonic $D$ decays. We remove these backgrounds by rejecting the signal $B$ candidates where two of their daughters are kinematically compatible with originating from a charm decay. After this requirement, we reject the continuum background using a cut on the likelihood ratio $R$, based on particle identification and event shape variables.

\begin{figure}[!htb]
     \begin{center}
     \includegraphics[width=0.4\linewidth]{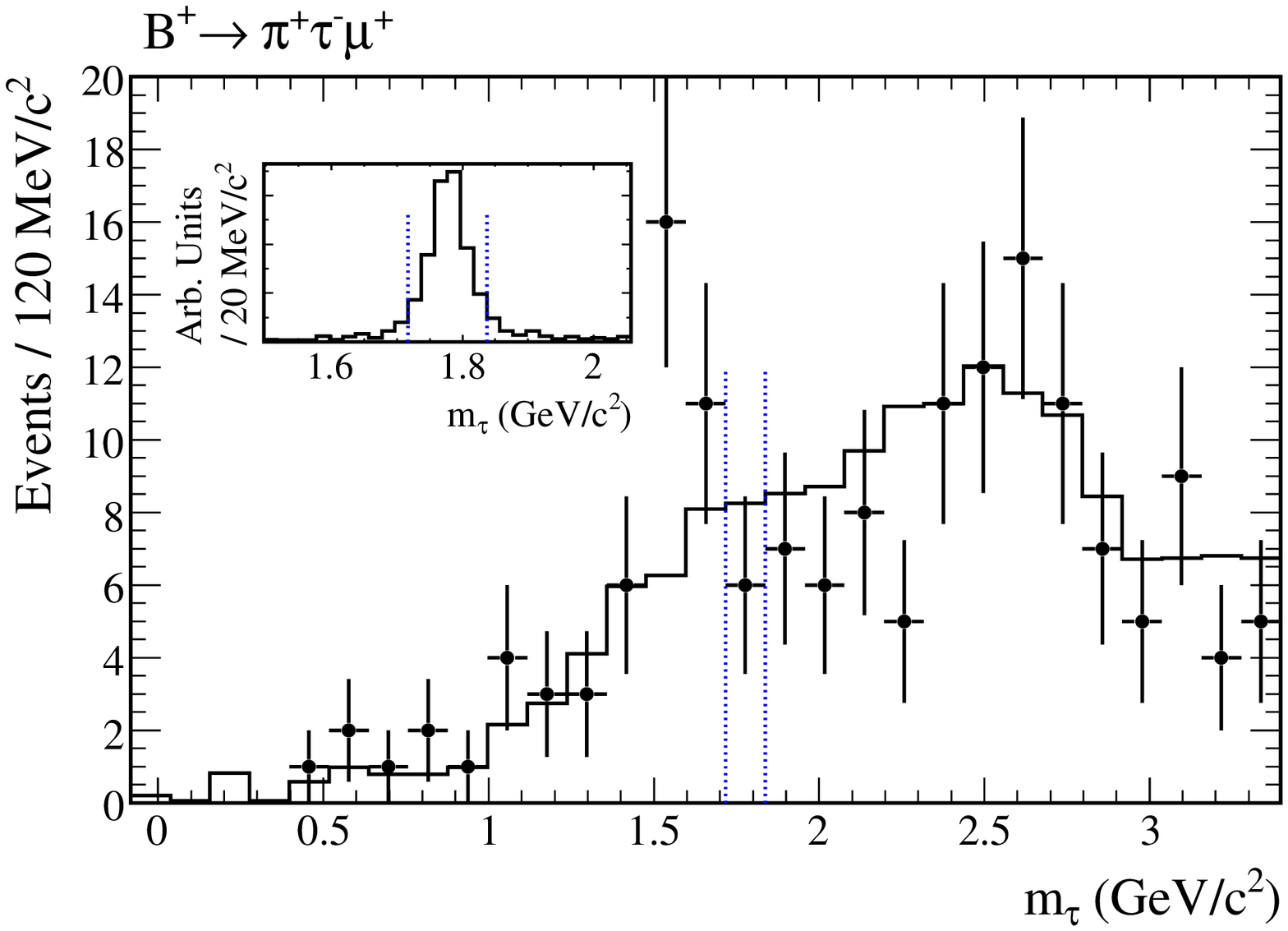}
     \includegraphics[width=0.4\linewidth]{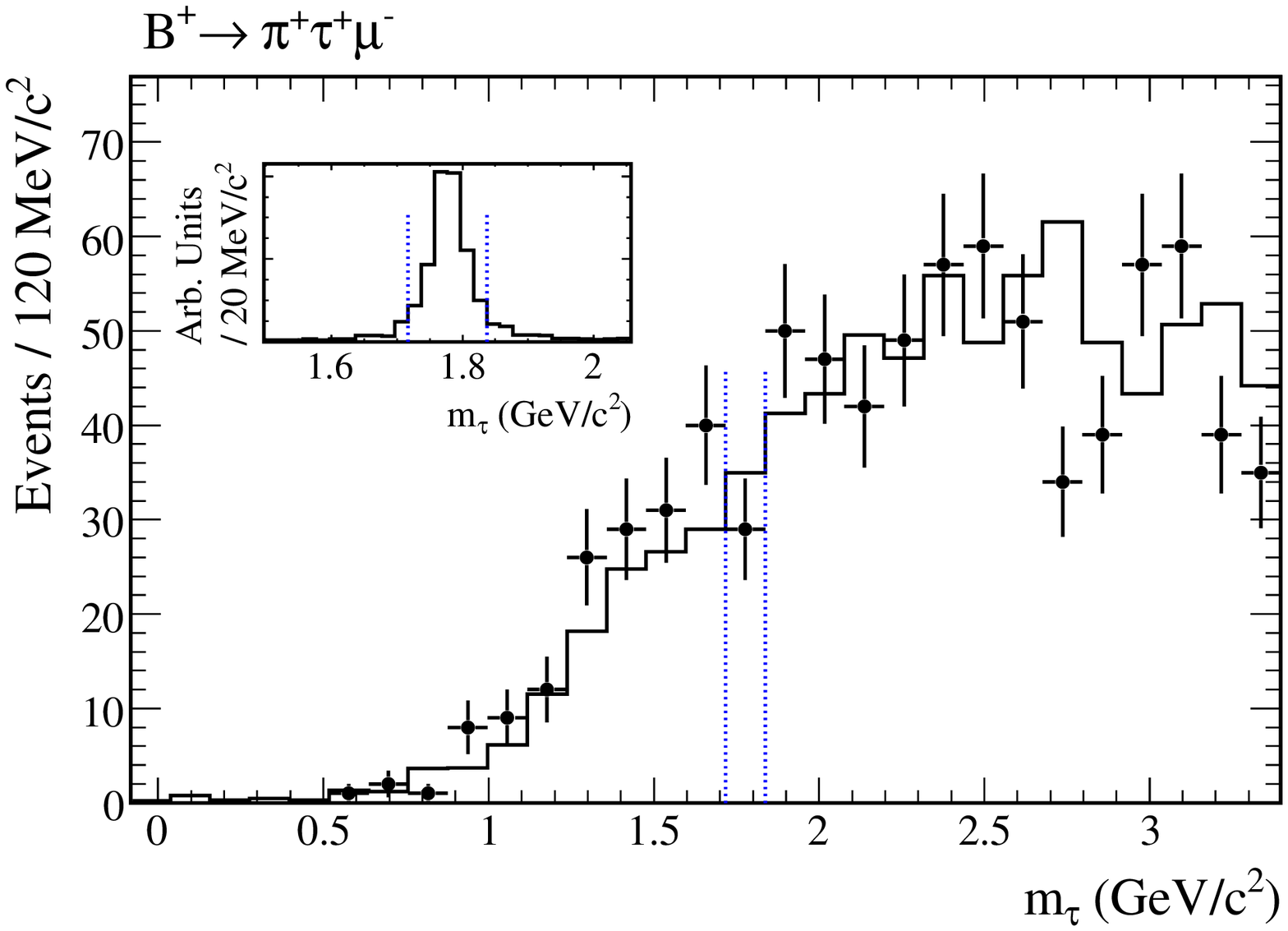}
     \includegraphics[width=0.4\linewidth]{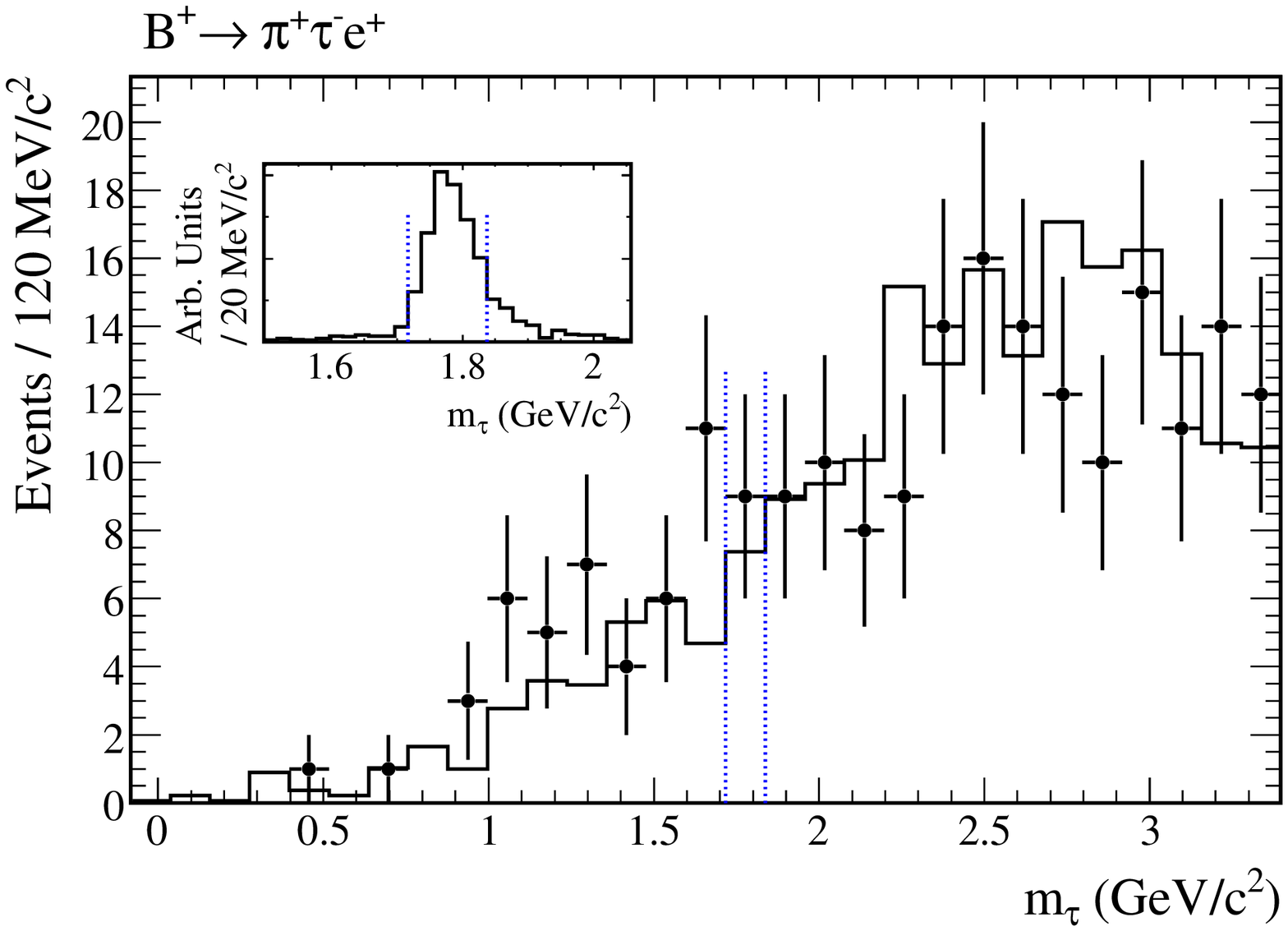}
     \includegraphics[width=0.4\linewidth]{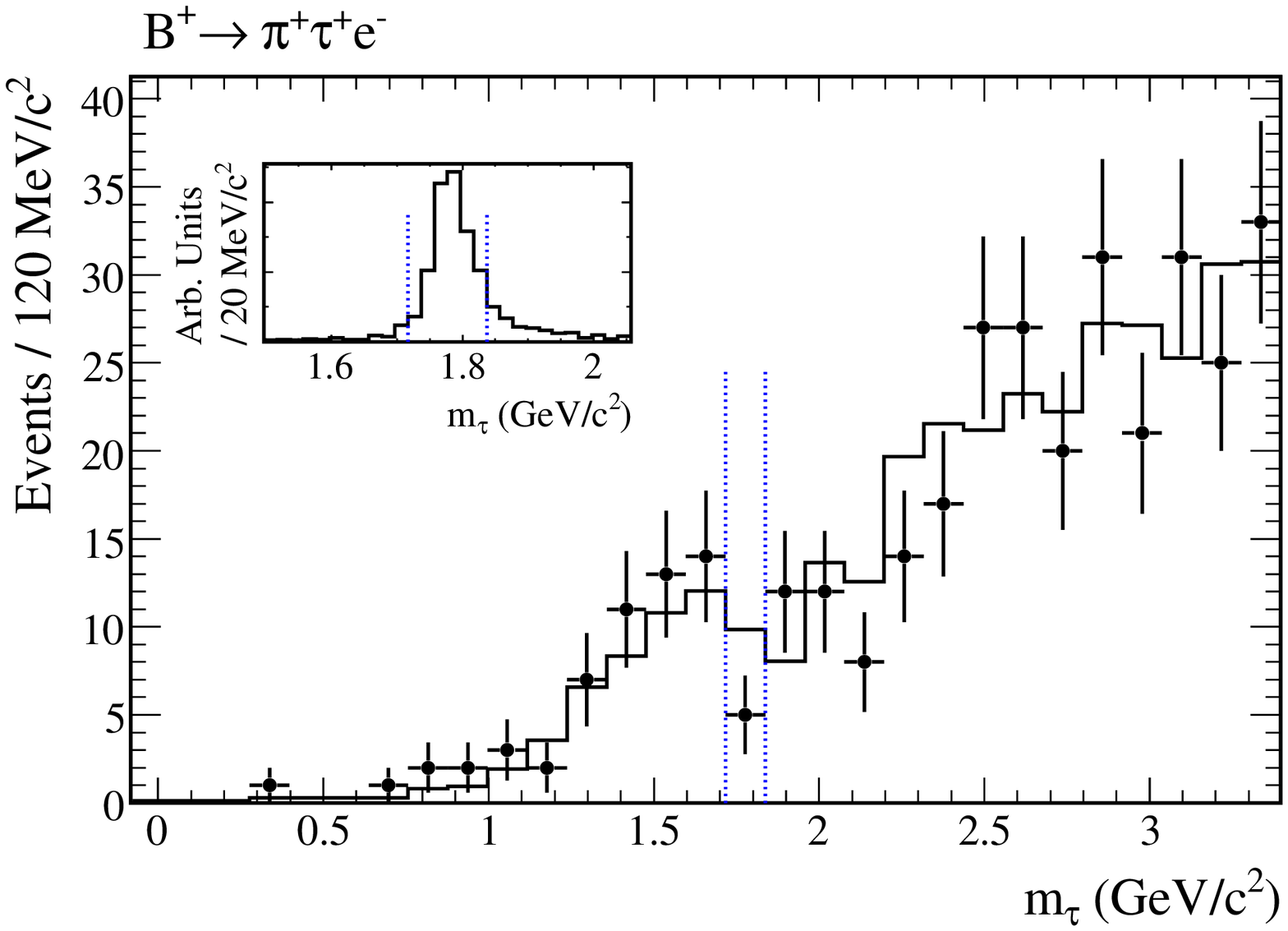}
     \caption{
         Observed distributions of the $\tau$ invariant mass for the \btoptl\ modes.
         The distributions show the sum of the three $\tau$ channels ($e$, $\mu$, $\pi$).
         The points with error bars are the data.
         The solid line is the background MC which has been normalized to the
         area of the data distribution.
         The dashed vertical lines indicate the $m_\tau$ signal window range.
         The inset shows the $m_\tau$ distribution for signal MC.
     }
     \label{fig:mtau-pitaul}
     \end{center}
   \end{figure}

The signal region is defined as $\pm 60 \mevcc$ around the indirectly reconstructed $\tau$ mass $m_\tau$. Figure~\ref{fig:mtau-pitaul} show an example of $m_\tau$ distributions for the data, for the background, and for the signal MC. No signal is observed, which allows us to put 90\% CL limit on the branching fractions. Assuming ${\cal B}(B^+ \to h^+ \tau^- \ell^+) = {\cal B}(B^+ \to h^+ \tau^+ \ell^-)$, we obtain the combined limits shown in Table~\ref{tab:combined-6chan-limits}. These limits can be translated into model-independent bounds on the energy scale of new physics in flavor-changing operators~\cite{ref:htlscale}: $\Lambda_{\bar b d}>11$~TeV and $\Lambda_{\bar b s}>15$~TeV (at 90\% CL), which improved the previous limits of 2.2 and 2.6~TeV, respectively.

  \begin{table}[!htb]
     \begin{center}
       \caption{
          Branching fraction central values and 90\% CL upper limits (UL)
          for the combination
          ${\cal B}(B^+ \to h^+ \tau \ell) \equiv {\cal B}(B^+ \to h^+ \tau^- \ell^+) + {\cal B}(B^+ \to h^+ \tau^+ \ell^-)$.
       }
       \label{tab:combined-6chan-limits}
       \begin{tabular}{|l|c|c|}
         \hline
                      &  \multicolumn{2}{|c|}{ ${\cal B}(B \to h \tau \ell)$ $(\times 10^{-5})$} \\
          \ \ \ \ \  Mode \ \ \ \ \ \ \ \ \ \ \        &  \multicolumn{2}{|c|}{\ \ \ \ central value\ \ \ \  \ \ \ 90\% CL UL \ \ \ } \\
         \hline
         $B^+ \to K^+ \tau \mu$   &  ~~~~~~$0.0\ ^{+2.7}_{-1.4}~~~~~~$   &  $<4.8$   \\
         \hline
         $B^+ \to K^+ \tau e$     &  $-0.6\ ^{+1.7}_{-1.4}$  &  $<3.0$   \\
         \hline
         $B^+ \to \pi^+ \tau \mu$ &  $0.5\ ^{+3.8}_{-3.2}$  &  $<7.2$   \\
         \hline
         $B^+ \to \pi^+ \tau e$   &  $2.3\ ^{+2.8}_{-1.7}$  &  $<7.5$   \\
         \hline
       \end{tabular}
     \end{center}
   \end{table}

\section{Search for $CP$ violation in the decays $\tau^-\to\pi^-\KS\left(\geq 0\pi^0\right)\nut$}

$CP$ violation, until now, has only been observed in hadronic decays ($K$, $B$, and $D$ systems). However, Bigi and Sanda predict~\cite{ref:bs} a non-zero decay rate asymmetry for $\tau$ decays to final states containing a $\KS$ meson, due to the $CP$ violation in the kaon sector. The decay rate asymmetry is:
\begin{equation*}
{\asy} = \frac{\Gamma\left({\taupksb}\right) - \Gamma\left({\taupks}\right)}
              {\Gamma\left({\taupksb}\right) + \Gamma\left({\taupks}\right)}, 
\end{equation*}
and is predicted to be equal to $(0.33\pm0.01)\%$. Any deviation from the SM prediction could be a sign of new physics. It has to be noted that $\asy$ is independent of the number of neutral pions in the final state. Recently, Grossman and Nir~\cite{ref:GNeff} noticed that the calculation needs to take into account interferences between the amplitudes of intermediate \KS and \KL mesons, which are as important as the pure \KS amplitude. This means that $\asy$ depends on the reconstruction efficiency as a function of the $\KS \rightarrow \pi^+\pi^-$ decay time.

We study here~\cite{ref:cpv} the decay channel $\tau^-\to\pi^-\KS\left(\geq 0\pi^0\right)\nut$. The event is divided into two hemispheres, one corresponding to the signal side, and one to the tag side with $\tau^- \to \ell^- \overline{\nu}_\ell \nut$, $\ell=e,\mu$. The selection of the signal events requires that the invariant mass of the reconstructed $\tau$ lepton is smaller than $1.8 \gevcc$.
After a two-stage likelihood selection (reducing mainly continuum background and \KS one, respectively), we obtain $199064$ candidates for the electron tag channel ($e$-tag), and $140602$ for the muon tag channel ($\mu$-tag). The composition of the sample in term of signal and background events is presented in Table~\ref{tab:cpv}.

\begin{table}[!htb]
\caption{Composition of the sample after all selection criteria have been applied.}
\label{tab:cpv}
\begin{center}
\begin{tabular}{|l|c|c|}
\hline
Source                         & \multicolumn{2}{|c|}{Fractions (\%) } \\
                               & \multicolumn{2}{|c|}{\hspace{0.5cm} $e$-tag \hspace{0.5cm} \hspace{0.5cm} $\mu$-tag \hspace{0.5cm}} \\
\hline
$\tau^- \rightarrow \pi^- \,\KS (\geq 0\pi^0) \, \nut$  & $78.7 \pm 4.0$  & $78.4 \pm 4.0$  \\
$\tau^- \rightarrow K^-  \,\KS (\geq 0\pi^0) \, \nut$  & $4.2 \pm 0.3$   & $4.1 \pm 0.3$   \\
$\tau^- \rightarrow \pi^- \,K^0 \overline{K}^0 \, \nut$          & $15.7 \pm 3.7$  & $15.9 \pm 3.7$  \\
Other background      & $1.40 \pm 0.06$ & $1.55 \pm 0.07$  \\
\hline
\end{tabular}
\end{center}
\end{table}

We need to correct the raw asymmetry from the pollution of the other modes shown in Table~\ref{tab:cpv}. For the mode $\tau^- \rightarrow K^-  \,\KS (\geq 0\pi^0) \, \nut$, the expected asymmetry is opposite to the one from the signal, and for the mode $\tau^- \rightarrow \pi^- \,K^0 \overline{K}^0 \, \nut$, the expected asymmetry is zero. Furthermore, an additional correction was pointed out recently~\cite{ref:ko}: we need to take into account a correction on the asymmetry $\asy$ due to the different nuclear-interaction cross-section of the $K^0$ and $\overline{K}^0$ mesons with the material in the detector. We calculate this correction to be $(0.07\pm0.01)\%$ and we subtract it from the measured asymmetry. After all corrections are applied, and after combining the results from the $e$-tag and $\mu$-tag, we obtain $\asy=(-0.36\pm 0.23\pm 0.11)\%$.

As we mentioned, this result should be compared with the prediction of the SM, corrected by the $\KS \rightarrow \pi^+\pi^-$ decay time dependence. Using the MC sample, we find a multiplicative factor of $1.08\pm0.01$. The SM decay-rate asymmetry, after correction, is then predicted to be $\asy^{\mathrm{SM}}=(0.36\pm0.01)\%$. We observe that our measurement is 2.8 standard deviations away from the SM prediction.




{\raggedright
\begin{footnotesize}



\end{footnotesize}
}


\end{document}